\documentclass[showpacs,twocolumn,eqsecnum,amsmath,final,prl]{revtex4}

\usepackage{amssymb}
\usepackage{bm}
\usepackage{array}
\usepackage{dcolumn}
\usepackage{graphicx}
\usepackage{times}

\renewcommand{\newblock}{}

\begin{document}
\title{High resolution strain measurements in highly disordered materials}

\author{Mark Sutton}
\affiliation{Physics Department, McGill University, Montr\'{e}al, Quebec H3A 2T8, Canada}
\author{Julien R.M.  Lhermitte}
\affiliation{Center for Functional Nanomaterials, Brookhaven National Laboratory, Upton,
New York 11973 US}

\author{Fran\c{c}oise Ehrburger-Dolle}
\affiliation{Univ. Grenoble Alpes, CNRS, LIPhy, F-38000, Grenoble, France}

\author{Fr\'ed\'eric Livet}
\affiliation{SIMaP, Grenoble INP-CNRS-UJF, BP. 75, 38402 Saint Martin d’H\`eres cedex, France}

\begin{abstract}
	The ability to measure small deformations or strains is useful
for understanding many aspects of materials.
Here, a new analysis of speckle X-ray diffraction peaks
is presented in which the systematic shifts of the speckles
are analyzed allowing for strain (or flow) patterns
to be inferred.
This speckle tracking technique measures strain patterns with a accuracy similar
to X-ray single crystal measurements but in amorphous or highly disordered
materials.
\end{abstract}
\maketitle

The measurement of small deformations in samples is a useful method to
characterize the properties of materials, soft materials in particular.
The properties of soft condensed matter systems are often controlled
by small mesoscale structural changes and/or deformations that have energy
scales comparable to thermal energies. These changes are the origin of many of
their visco-elastic properties.
Deformations are typically measured via strain, or the geometrical
relative displacements of elements in a body. Small deformations
are hard to measure. High resolution strain has
been measured using X-rays and neutron scattering
(\cite{robinson2009coherent}, \cite{allen1985neutron},
\cite{nielsen2003measurements}) and visible light
(\cite{chu1985applications}, \cite{sutton1983determination},
\cite{yamaguchi1981laser}).  However, these measurements are mostly
limited to crystalline materials and/or macroscopic sections of samples.
There are few measurement techniques involving
amorphous materials with X-rays. Nielsen
et al ~\cite{nielsen2003measurements} applied a 3D
absorption tomography technique. This technique provides a high
resolution 3 dimensional picture of the static strain of materials.
However, this technique requires measuring many diffraction patterns,
potentially restricting its use for
in-situ measurements.
In this paper an extension of X-ray photon correlation spectroscopy (XPCS)
is proposed for in-situ measurements of strain in amorphous materials.
It measures a 2D projection of the strain but the
technique only requires taking two coherent diffraction patterns.
Since the analysis is somewhat involved, the emphasis of this paper is
simply on the technique itself.

For a simple insight into the technique,
imagine a deformation of the underlying space,
$\vec{r}^{~\prime} = \mathrm{M}\cdot\vec{r}$.
For small deformations, the transformation consists of a possible
rigid body displacement and a local distortion.
From this it follows that the strain is the difference between $M$
and the identity matrix and related to the gradient of $M$ for small
deformations.
That is $\Gamma_s = M - I$ is the strain
tensor.
When $M$ is independent of $\vec{r}$ a homogeneous deformation
results.
Since diffraction is measured in reciprocal space, we consider
the effect on the density, $\rho(\vec{r})$, and its
Fourier transform, $\rho(\vec{q})$. To get the Fourier transform
of the deformed material, $\rho^\prime(\vec{q})$:
\begin{eqnarray}
\rho^{\prime}(\vec{q})
    &=&  \int \rho(\mathrm{M}\cdot\vec{r}) e^{-i\vec{q}\cdot\vec{r}} d\vec{r}\nonumber\\
    &=& \rho(\mathrm{M}^{-T}\cdot\vec{q})/ \mathrm{det}(M),
\end{eqnarray}
by simple substitution.
This calculation assumes a negligible change in the volume of integration.
Speckle measurements use a small diffraction volume and this sets the
scale determining when a deformation is small and for which the
approximation is to be valid.
Another approach is given by solving
the convection-advection diffusion equation by the method of characteristics
and may be found in Fuller et. al~\cite{fuller1980measurement}.
So, if we can relate features in $\rho(\vec{q})$ (or scattering intensity)
before and after the deformation, their time dependent shifts in reciprocal space are:
\begin{equation}
	\Delta \vec{q} = M^{-T}\cdot \vec{q}-\vec{q}= -\Gamma_s^T\cdot\vec{q}
\end{equation}
where $M^{-T}\approx I-\Gamma_s^T$ for small deformations.
For a simple velocity field
\begin{eqnarray}
	M(t)\cdot\vec{r} &=&  \vec{r}+\int_0^t \vec{v}(\vec{r},\tau)d\tau\nonumber\\
	&=& \vec{r}+(\vec{v}_0+\Gamma\cdot\vec{r})t \nonumber\\
	&=& \left( I+\Gamma_s\right)\cdot \vec{r}.
\end{eqnarray}
showing the strain tensor is related to the velocity gradient matrix,
$\Gamma = \nabla \vec{v}(\vec{r})$
for a velocity field.
The rigid body shift gives a phase factor, $\exp(i\vec{v}_0\cdot\vec{q}\tau )$,
due to the shift in $\vec{r}$ and will be
ignored as it will not be seen in the scattered intensity. It is
also assumed that the velocity field is in steady state for the time
of the measurement of each diffraction pattern.

Using these ideas for the analysis of diffraction patterns involves,
relating features in the starting diffraction pattern and the deformed
diffraction pattern, measuring the shift in these wave-vectors and analyzing
the shifts to obtain the strain matrix.

The illumination of a disordered material with a coherent source leads to
a complicated interference pattern modulating the conventional
(incoherent) diffraction pattern. This
pattern is typically called a speckle pattern. The pattern 
reflects more than one length scale. There is a set of short length scales
which reflect the scattering of the pieces or blocks which lead to
longer distances in the diffraction pattern. These are also reflected
in conventional
X-ray diffraction and their analysis is well developed. The speckles
have short length scales in reciprocal space and arise due to the finite
size of the beam. The rule of thumb is that the modulating speckle pattern
has many peaks of widths $2\pi/L$ where $L$ is the
appropriate beam dimension on
the sample. At each point in the diffraction pattern there is a contribution
from each of the many blocks or subsections each with its own phase factor
randomly varying from 0 to $2\pi$.
These phases will add up to a well defined phase under coherent illumination.
If they add in phase
the result is constructive interference and the intensity is high. Where
they add up out of phase, there is low intensity. Each speckle peak is
somewhat like a local Bragg peak.
Below we show that tracking the positions of speckles can be
used to infer small deformations.

The shift in wave-vectors after a deformation can be obtained
from the following correlation function:
\begin{equation}
	\label{eqng2}
        \mathrm{g}_2(\vec{q}_0, \Delta \vec{q}, \tau) = \frac{\langle
I(\vec{q},t) I(\vec{q}+\Delta \vec{q},t+\tau) \rangle_{\vec{q}}}{\bar{I}^2}.
\end{equation}
Since each wave-vector shifts by different amounts the average over
$\vec{q}$ is over a 
small region centered around $\vec{q}_0$. This assumes that the 
displacement is slowly varying in $\vec{q}$.
Notice that when the shifts are time invariant or slow, it is possible to
also average in time.
When
the scattering is sharply peaked, as for Bragg peaks and for speckles,
one can measure $\Delta \vec{q}$ by following the local maxima.
For Bragg peaks one measures the shifts of a few peaks. For
speckles, the cross-correlation measures the average shift of the speckles
over the region averaged. As pointed out below this is over many
hundreds of speckles.
The correlation function, $\mathrm{g}_2(\vec{q}_0, \Delta \vec{q}, \tau)$,
is an extension of the intensity-intensity
correlation function used in XPCS. The time correlations for XPCS
are calculated using $\Delta \vec{q} = 0$.

The experiments were carried out at beamline 8-ID-I of the Advanced
	Photon Source. For this setup, the energy was
	7.488~keV($\lambda=1.663$~\AA)
	monochromated by double bounce Ge(111) crystals.
	The incident flux was $\approx 10^{9}$ photons/sec through
	a 20$\times$20~$\mu$m$^2$ aperture. 
A direct-illumination
deep-depletion CCD (PI 1152 $\times$ 1242, 22.5 $\mu$m resolution) was
used as an area detector.
	Each pixel in the area detector corresponds to
	$2.0\times 10^{-5}$~\AA$^{-1}$, which is close to the speckle
	size. The filled rubber samples were held in a vacuum chamber with
	an in-situ tensile stress-strain cell. The exposure time per frame was
	0.1 second recurring every 2.0 seconds.
	Further details are given in
	references~\cite{dolle01,dolle02}.

The sample consists of a cross-linked elastomer (Ethylene Propylene Diene
Monomer, EPDM, rubber) filled with hydoxylated
pyrogenic silica (Aerosil 200~\cite{rieker01,dolle03,dolle04}).
The volume fraction of silica is close to 0.16.
In our measurements, the
one millimeter thick sheets are punched out to a classical dumb-bell
shape (width = 4 mm).
As described in Ref.~\cite{dolle01} upon applying a step strain on the sample,
the stress jumps and then slowly relaxes as the sample is help at constant strain.
For this article, only one 400 second data set is presented for a sample which
was stretched by 60\% and after which the macroscopic strain is fixed.
The data collection for this run started approximately
1250 seconds after the application of the 60\% strain step.
This data is part of those presented in Ref.~\cite{dolle01} where emphasis
is on the understanding of the underlying science of elastomers.
Here, since the analysis is somewhat involved, this article emphasizes
the new data analysis.
Only a single run is described as this simplifies the discussion
while still demonstrating the
generality of the technique. The technique measures small strains in
disordered materials and does not only apply to the elastomers used here.
Using this analysis on the other data sets presented in Ref.~\cite{dolle01}
confirms the ideas presented here and should be used to study the
viscoelastic properties of the elastomers which were studied.

The 200 frames of the data set have been sequentially averaged by 5 frames
to produce 40 images which are analyzed below.
To select the small regions of $\vec{q}$ for the cross-correlations,
the scattering images are decomposed into small wedges or bins of $|q|$
and $\phi$.
	The orientation of the detector is such
	that vertical on the detector is vertical on the sample.
	The azimuthal angle $\phi$ is
	$\mathrm{atan}(q_{\mathrm{vert}}/q_{\mathrm{hor}})$.
	The wedges
	are 20 pixels (0.00040~\AA$^{-1} $) wide in $|\vec{q}|$ and 10$^\circ$ in
	$\phi$. Cross-correlations for all wedges with more
	than 1000 pixels
	up to wave-vector 0.024~\AA$^{-1}$ (1200 pixels) were calculated.
	This gave 376 non-overlapping wedges. Bins are numbered with $\phi$ increasing for
	fixed $q$ and then $q$ increases for the next set of $\phi$.
	Figure 1 shows a typical bin,
	(bin 57, $q$=0.0090~\AA$^{-1}$, $\phi$ = 200$^\circ$)
	and its cross-correlation between the first averaged frames.
	As for each cross-correlation~\cite{footnote1,padfield2012}, it has a
	peak amplitude of $1+\beta$ (speckle contrast)
	sitting on a background of one. Each cross-correlation is least
	squares fit to a 2D Gaussian peak. Peak widths give
	the speckle size. The shift of the peak maximum gives
	the speckle movement.
	The cross-correlation is calculated for integral pixel shifts,
	with no shift defining zero and points on each side being plus
	or minus a one pixel shift. If there is no peak shift the cross-correlation
	would be symmetric about zero.
	It is easy to see if a shift by a fraction of a pixel exists
	as this leads to asymmetric values of the data at the $\pm 1$
	positions as shown in Fig.~1c.
	Figure~1c also
 shows sections of the 2D Gaussian fit for this cross-correlation. Note for
this example the center of the peak is offset by (0.24,-0.10) pixels or
$(4.8\times 10^{-6},-2.0\times 10^{-6})$~\AA$^{-1}$ in $(q_{hor},q_{vert})$
from the center or the unshifted correlation point. The full width at
half maximums are 2.733 and 2.093 pixels in the horizontal and vertical
directions and $\beta=.263$.
	A preliminary version of this
	analysis is given in Lhermitte's PhD thesis~\cite{lhermitte2011using}.

Figure~2 shows the intensity of the time averaged
small angle scattering (the blank rectangle in the
upper right is the beam stop). 
Superimposed is a representation of the speckle shifts between the first
image and two
other times.
The shift for each bin is plotted as an arrow
starting at the center of the bin.
Since the shifts are so small, the shift in pixels 
has been multiplied by 10 where one pixel is $2.0\times10^{-5}$~\AA$^{-1}$.
One can immediately see that the relaxation
of the filled rubber has a hyperbolic flow pattern. It is away from the beam
center in vertical ($\phi=270^\circ$) which is along the tensile strain
and towards the beam center in
the horizontal ($\phi=180^\circ$). One also sees the shifts increase with
$|\vec{q}|$. For reference, bin 57 is plotted in red.

Equation~\ref{eqng2} is similar to the equation for XPCS except the analysis is
only for the $\Delta \vec{q}=0$ term.
XPCS has demonstrated its ability to measure time constants from
microseconds to many hours. It has demonstrated that static disorder
leads to speckle patterns unchanging in time, also for hours.
Figure~3 shows the $g_2(\tau)$, 
as one would calculate it in a XPCS measurement.
The slower decaying solid line in Fig.~3 is obtained
by following the peak maximum from fitting the cross-correlation
of the first averaged by 5 patterns in time versus each consecutive averaged by
5 sequence of patterns~\cite{footnote2}.
The decay time of the slower curve
is the average time it takes for speckles in bin 57 to move through the
Ewald sphere in three
dimensional reciprocal
space. To demonstrate stability, data points
for $g_2$ from a silica aerogel and a packed Aerosil 200 silica
(Ref.~\cite{dolle01}, Fig.~7a)
measured with
the same setup are shown.
They do not 
decay for the times shown. The cross-correlation analysis of the aerogel
data shows
the speckles do not shift in $\vec{q}$. This can also be inferred from
the figure as any shift would cause $g_2$ to decay.

The fastest decaying curve in Fig.~3 is an example of the type of curves
analyzed in Ref.~\cite{dolle01} using 
conventional relaxation correlation functions.
Figure~1c shows that speckles are approximately 2 pixels wide. Figure~4
shows that for about 100 seconds and longer the speckles for bin 57
have shifted by more than 2 pixels. This is origin of the loss of correlation
for time delays longer than 100 seconds.
This new analysis shows that the decay in time for $g_2$
is due to the shift of the speckles and not to random diffusion.
That the shifts increase linearly in $\vec{q}$ explains why the
time constants vary as $1/q$.
It also shows that the time constants measured in Ref.~\cite{dolle01} are
indirect measures of velocity gradients.
It is important to realize that the motion of the speckles is happening at
constant macroscopic strain. The measured motion is accompanied by a changing
macroscopic stress.
The speckle motion is expected in all three dimensions.
We expect the time for the speckle to move perpendicular to the detector plane
to be comparable for the time of in-plane motion given the.
uniaxial nature of the applied strain,
The small angle X-ray scattering (SAXS) with the speckle averaged away is equivalent to
a conventional SAXS measurement. The SAXS pattern is isotropic and did not
vary for the 200 images of this data set. 

A cross-correlation can be calculated for
each bin and any pair of diffraction patterns. Only the shifts measured for
each subsequent pair of
patterns are used here. Single frame
cross-correlations give quite acceptable correlations but for this analysis
5 images are averaged and then correlated as this gives cleaner
images. Since the correlations die away in minutes,
the shifts are measured by cross-correlating each nearest time pair of the
averaged scattering. Using patterns further separated in time leads to
compatible shifts, but have weaker correlations and are less accurate.
Since the deformations are varying slowly across the
whole run, it is better to measure nearest neighbour shifts and integrate
the shifts for longer time intervals. Summing the shifts leads to
a cumulative shift over the run and is shown in Fig.~4
for bin 57 ($q$=0.0090~\AA$^{-1}$, $\phi$ = 200$^\circ$).

For a given time delay and from the cumulative shifts in each bin, a
vector field may be calculated.
Two examples of these are  plotted in Fig.~2.
For each delay time
slice, the vector field can be fit to a velocity gradient:
$d\vec{q}/dt=-\Gamma\cdot \vec{q}$
where $d\vec{q}$ comes from the measured shifts and $dt$ is the time
between images.
Each vector field is well described by a diagonal 2D matrix.
Since it is difficult to compare 2D vector field images,
the quality of a typical fit is shown in Fig.~5 by a comparison between
the fit and the measured vector components for the first 350 bins.
This gives a quantitative representation of the quality of the fit.
Since the fits agree well with the data, the white arrows on the right
side of Fig.~2
also shows this vector field in another representation.
Explicitly for this
image the deformation is: 
$(\Delta q_{hor},\Delta q_{vert})= (-0.0089 q_{hor}, .0106 q_{vert})$
and the ratio $\Delta \vec{q}/\vec{q}$ gives the diagonal elements of the
strain tensor.
A simple two
parameter fit to all shifts of a given time delay fits all $q$-$\phi$ bins exceptionally well.

Figure~6 shows the evolution of the diagonal elements of
the velocity gradient tensor $\Gamma$ as a function
of time obtained from all vector field fits. Integrating this over
time gives the strain as discussed above. Remember that here
$\Gamma$ is a velocity
gradient tensor. For reference, 
$\Gamma_{\mathrm{vert}}=35.0\times 10^{-6}~s^{-1}$ 
corresponds to points
1~$\mu m$ apart in the vertical direction having a
difference in vertical velocity of $35$~\AA/sec.
These velocity gradients are quite compatible with the
coarser ``ballistic'' velocity estimates using a compressed
exponential correlation function based on
Cipelletti et al~\cite{cipelletti} (see Ehrburger-Dolle et al
with an earlier discussion of this data~\cite{dolle01}).
The evolution of $\Gamma$ with time
reflects the slow visco-elastic relaxation in the
silica filled rubber while the sample is held at constant macroscopic
strain.

In conclusion, it has been shown that a simple extension of XPCS can
measure the projection of the strain tensor across the scattering volume.
Since the dimensions of the beam are $20\times20~\mu$m$^2$ this gives a submicron
measurement of the local strain fields and the strain precision is similar to
what can be measured in a single crystal measurement using the shifts in a Bragg
peaks. It is stressed here that this analysis works for amorphous or
highly disordered materials, in particular most heterogeneous
soft matter systems.
Details of the rheological implications for our samples and
for the conventional XPCS analysis~\cite{dolle07} from this new
approach is left for future articles. It is worth pointing out that
the decay of the XPCS signal in this system can be explained by the
visco-elastic flow of the filled rubber, without any random
Brownian motion.

We emphasize that the main result of this paper is that small strain fields
in amorphous materials
can be measured by cross-correlating speckle patterns from before and after
a change in strain.
Nothing in the above analysis is specific to elastomers except maybe for
the uniaxial approximation used to fit the vector field. Other distortions
may require a different model. We emphasize, that
the strain measurement only requires two images with sufficient intensity
to measure their speckle patterns and that the two images are separated in
time by less then the time it takes for the speckles to move off the
Ewald sphere~\cite{footnote3}.
Too large a change in strain during exposure time of the images will
smear the speckles and reduce sensitivity. Also, if the out of plane
strain between
the two images is too large there will be no correlation and the
strain can not be measured. For SAXS, movement
along the beam in either the forward or backward direction have no
in-plane component and so does not contribute to the diffraction pattern,
but it can contribute to the motion of the speckle peaks. Also
in-plane motions are averaged through the sample along the beam.

The advent of the new lattice structures that are being used
to upgrade the X-ray synchrotrons
will lead to an increase in
coherence by two to three orders of magnitude.
Also a new generations of X-ray area detectors with multi kHz rates
are becoming commercially available and still faster ones are in
developement.
Together this will enable speckle measurements to be easily performed with
times on order of microseconds. For this set of experiments it will
allow for measuring strains during
the initial stage of deformation when the macroscopic strain is applied.
We also point out
that one of the nice features of this new technique is one can
easily measure nanometer sized distortions and even slow
measurements of these are extremely important for creep and aging
studies in materials.

\begin{acknowledgments}
This research used resources of the Advanced Photon Source, a U.S.
Department of Energy (DOE) Office of Science User Facility operated for
the DOE Office of Science by Argonne National Laboratory under Contract
No. DE-AC02-06CH11357.
We thank the staff of beamline 8-ID-I for their excellent help.
\end{acknowledgments}

\newpage
\newpage


\begin{figure}
	\includegraphics[width=4in]{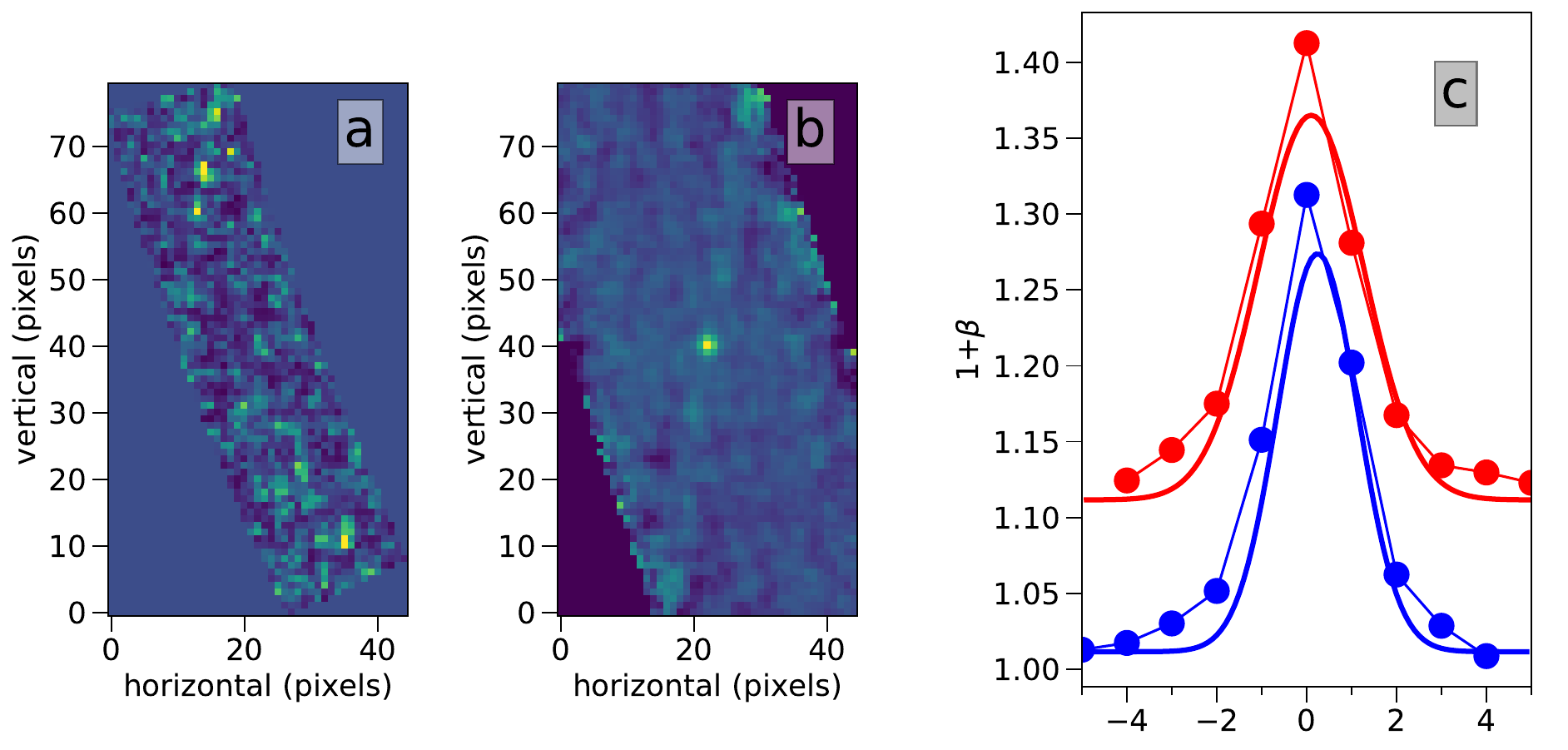}
	\caption{a) Speckle intensity for bin 57, $q$=0.009~\AA$^{-1}$,
	$\phi = 200^\circ$ as placed in a rectangular image.
	The underlying diffraction intensity
	is constant over the wedge 
	and the fractional intensity fluctuates by 
	approximately $\pm\sqrt{.3}=.55$ reflecting the speckle variation.
b) Cross-correlation between the first two averaged by 5 images.
	Scale can be determined from part c).
c) Line-outs along around the central pixel of the cross-correlation in the
vertical (red, offset in y by 0.1) and horizontal (blue) directions.
	The thick lines are for the 2D Gaussian that result from
	fitting the peak. The resulting shifts are 0.24 pixels in
	the horizontal and -0.10 pixels in the vertical and are reflected
	in the asymmetric placement of the measured points.}
\end{figure}


\begin{figure}
	\includegraphics[width=4in]{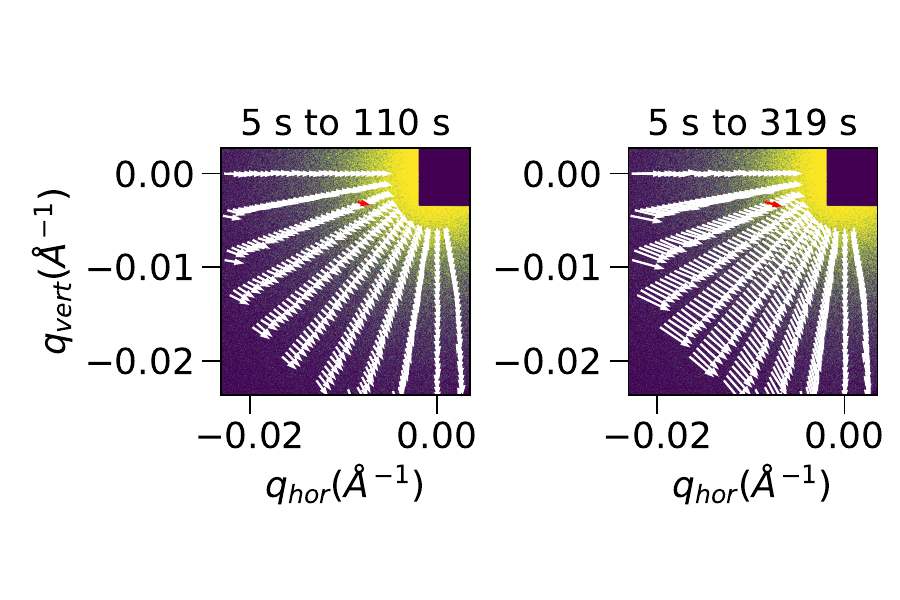}
	\caption{Strain field for images from times  5 to 110 seconds and
	for images at 5 to 319 seconds. Each local q region has a
	shift in the speckle pattern given by the arrows which are
	scaled up by 10 ($q$ per pixel = $2.00\times 10^{-5}$\AA$^{-1}$).
	The coordinates of the shifts for the right panel
	are plotted in Fig.~5. The shifts are superimposed on the underlying
	small angle X-ray scattering intensity and the blank rectangle in the
	upper right corner is from the beam stop. For reference, bin 57's shift
	is shown in red.}
\end{figure}



\begin{figure}
	\includegraphics[width=4in]{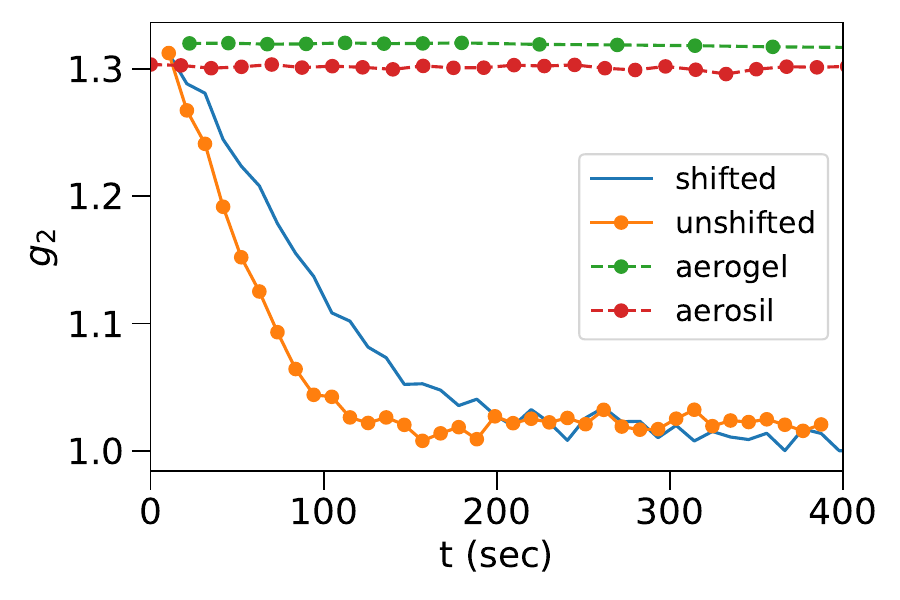}
	\caption{A conventional calculation of $g_2$ is shown by the
	line with dots, labelled unshifted ($\Delta\vec{q}=0$).
	A measurement following the in-plane shift
	observed on the detector is given by the solid line, labelled
	shifted. For reference,
	the green data points shows the measured data from a static aerogel sample and
	the red points for a packed aerosil 200 sample(from Ref~\cite{dolle01} Fig. 7a).
	The data analyzed is from the same $q$-$\phi$ bin as in Fig.~1.}
\end{figure}

\begin{figure}
	\includegraphics[width=4in]{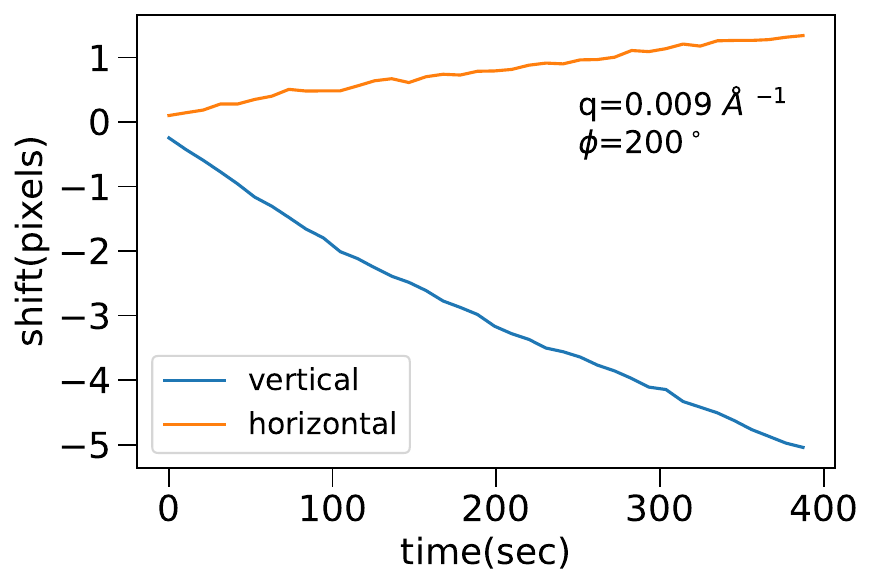}
\caption{The cumulative shift for bin 57 over the length
	of the time series. A pixel is $2.0\times 10^{-5}$~\AA$^{-1}$.}
\end{figure}

\begin{figure}
	\includegraphics[width=4in]{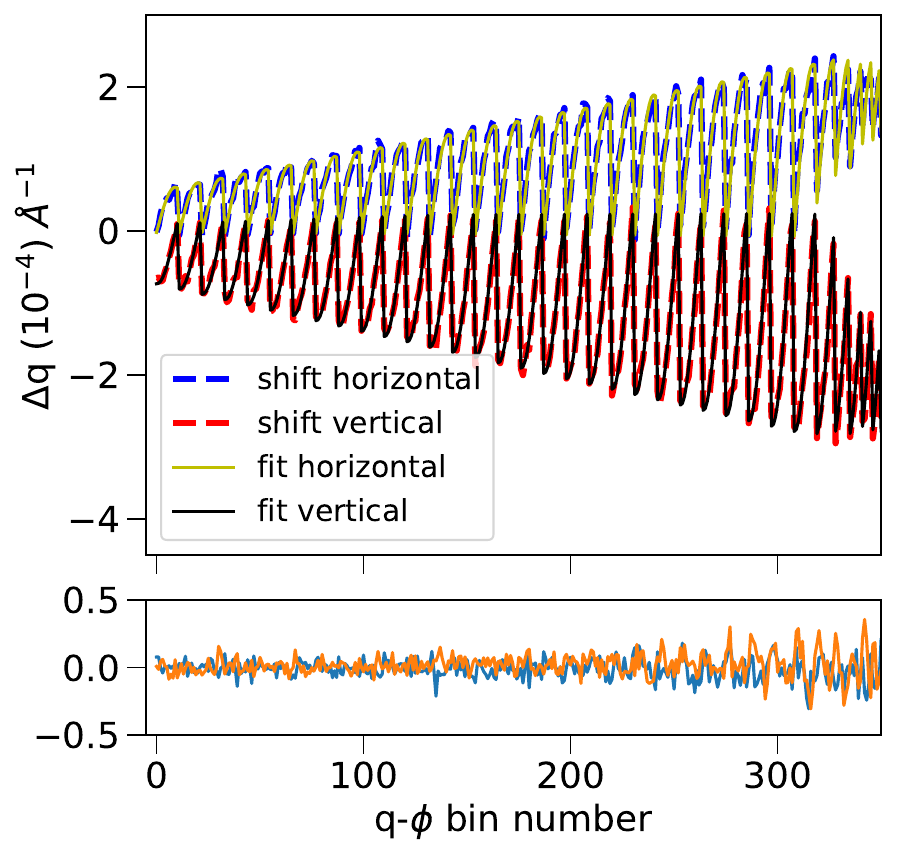}
\caption{Example of a two parameter fit to a simple diagonal
	deformation pattern. The two components of the vector shift
	are plotted separately, blue dashed for vertical and red dashed
	for horizontal. The fitted values (solid lines)
	overlap with the data and so the lower panel shows
	the two sets of differences between the fits and the measurements.
	The vector field used in the fit is the same as in the
	right panel
	of Fig.~2 and corresponds to $\Gamma_{vert}=31.97\times10 ^{-6}$ and
	$\Gamma_{hor}=-26.75\times 10^{-6}$ per second. These two parameters
	were used to generate the solid lines.
	The saw-tooth nature of the data reflects the order of the bins
	in the diffraction pattern. For a given $q$ the bins increase in
	$\phi$ and then bins increase in $q$ giving the next tooth.}
\end{figure}

\begin{figure}
	\includegraphics[width=4in]{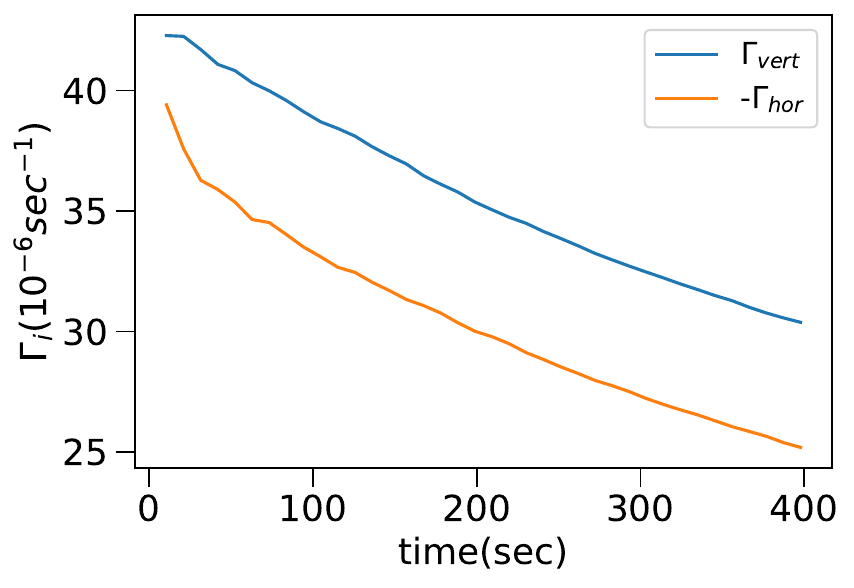}
	\caption{Time evolution of the velocity gradient parameters obtained by
	fitting the vector fields over the run. The cumulative strain is the
	integral of $\Gamma$ over time.}

\end{figure}


\begin{thebibliography}{}


\bibitem{robinson2009coherent}
I. Robinson and R. Harder,
		Nature materials, {\bf 8}, 291--298, (2009).

\bibitem{allen1985neutron}
A.J.~Allen, M.T.~Hutchings, C.G.~Windsor, and C.~Andreani,
		Advances in Physics, {\bf 34}, 445--473, (1985).

\bibitem{nielsen2003measurements}
S.F.~Nielsen, H.~F. Poulsen, F.~Beckmann, C.~Thorning, and J.A.~Wert,
		Acta Materialia, {\bf 51}, 2407--2415, (2003).

\bibitem{chu1985applications}
T.C.~Chu, W.F.~Ranson, and M.A.~Sutton,
Experimental mechanics, {\bf 25}, 232--244, (1985).

\bibitem{sutton1983determination}
M.A.~Sutton, W.J.~Wolters, W.H.~Peters, W.F.~Ranson, and S.R.~McNeill,
		Image and Vision Computing, {\bf 1}, 133--139, (1983).

\bibitem{yamaguchi1981laser}
I. Yamaguchi,
Journal of Physics E: Scientific Instruments, {\bf 14}, 1270, (1981).

\bibitem{fuller1980measurement}
G.G.~Fuller, J.M.~Rallison, R.L.~Schmidt, and L.G.~Leal.
Journal of Fluid Mechanics, {\bf 100}, 555--575, (1980).

\bibitem{dolle01} F. Ehrburger-Dolle,  I. Morfin, F. Bley, F. Livet,
G. Heinrich,  S. Richter, L. Pich\'e, and M. Sutton,
		Macromolecules, {\bf 45}, 8691-8701 (2012).
\bibitem{dolle02} F. Ehrburger-Dolle, I. Morfin, F. Bley, F. Livet,
	G. Heinrich, L. Pich\'e, and M. Sutton,
Journal of Polymer Science Part B: Polymer Physics {\bf 52}, 647-656, (2014).

\bibitem{rieker01} T.P. Rieker, M. Hindermann-Bischoff, F. Ehrburger-Dolle,
	Langmuir, {\bf 16}, 5588-5592, (2000).

\bibitem{dolle03} F. Ehrburger-Dolle, M. Hindermann-Bischoff, E. Geissler,
	C. Rochas, F. Bley, F. Livet, F.
Materials Research Society Symposium, {\bf 661}, KK7.4.1, (2001).

\bibitem{dolle04} F. Ehrburger-Dolle, F. Bley, E. Geissler, F. Livet,
	I. Morfin, C. Rochas,
Macromolecular Symposia, {\bf 200}, 157--167, (2003).


\bibitem{footnote1}
	Cross-correlations where calculated by taking
	each wedge and placing it in a rectangular matrix with a size
	of twice the tightest rectangle covering the wedge. Correlations where
	calculated using Fourier transforms by further extending the matrix to
	one with sizes that are multiples of 2,3 and 5.
	This insures the fast Fourier transform can be used.
	This rectangle has many zeros,
	including those outside the wedges and those due to bad
	or masked pixels.  To take these into account,
	let $B$ be a rectangular matrix of zeros and ones identifying which pixels
	have signal and $I$ be the matrix of the same size with the signal
	to be correlated.
	Then
	$g_2 = ((I_0\otimes I_1) \times (B\otimes B))/((B\otimes I_0) \times (I_1\otimes B))$
	where $\times$ is element by element multiplication~\cite{padfield2012}.
	The denominator amounts to symmetric normalization.

\bibitem{padfield2012}
D. Padfield,
IEEE Transactions on Image Processing, {\bf 21}, 2706-2718, (2012)

\bibitem{lhermitte2011using}
J. Lhermitte,
\newblock {\em Using Coherent Small Angle Xray Scattering to Measure Velocity
  Fields and Random Motion}.
		\newblock {PhD thesis, McGill University, (2011).}

\bibitem{footnote2} The speckle shifted contrast is the amplitude
	of the 2D Gaussian fit to the cross-correlation of the two
	images for the given time delay.

\bibitem{cipelletti}
	L. Cipelletti, L. Ramos, S. Manley, E. Pitard, D. A. Weitz,
E. E. Pashkovski and Marie Johansson,
Faraday Discuss., 123, 237–251 (2003).

\bibitem{dolle07}
F. Ehrburger-Dolle, I. Morfin, F. Bley, F. Livet, G. Heinrich, Y. Chushkin,
and M. Sutton,
Soft Matter {\bf 15}, 3796-3806 (2019).

\bibitem{footnote3} As long as there is partial contrast there will be speckles in the
	diffraction pattern and thus peaks. Being able to follow
	the shift of peaks
	in the diffraction pattern is all that is required for the analysis.
	The strain sensitivity will obviously be reduced for highly
	under sampled speckles as a larger shift in reciprocal space
	will be required to move the speckle with respect to the pixel
	size. It will also be more likely that a shift off the Ewald sphere
	will cause a loss in correlation.
	It thus helps if the intrinsic speckle size, which is determined
	by the diffraction volume, is comparable to the pixel
	size on the area detector. Speckles do not need to be
	over sampled.
\end{thebibliography}
\end{document}